\documentclass[final,5p,times,twocolumn,numbers,sort&compress]{elsarticle}

\usepackage[T1]{fontenc}
\usepackage{amsmath}
\usepackage{amssymb}
\usepackage{bm}
\usepackage{graphicx}
\usepackage{booktabs}
\usepackage{array}
\usepackage{microtype}
\usepackage{xcolor}

\newenvironment{revision}{\begingroup\color{red}\bfseries}{\endgroup}
\usepackage{ulem}
\DeclareRobustCommand{\erase}{\bgroup\markoverwith{\textcolor{red}{\rule[.5ex]{2pt}{0.4pt}}}\ULon}

\usepackage{placeins}

\journal{Physics Letters B}
\biboptions{sort&compress}

\newcommand{\includeorplaceholder}[2]{%
  \IfFileExists{#1}{\includegraphics[width=#2]{#1}}{%
    \fbox{\parbox[c][0.095\textheight][c]{#2}{\centering
      Figure file not supplied:\\[2pt]\texttt{\detokenize{#1}}}}%
  }%
}

\begin{document}

\begin{frontmatter}

\title{Charge-symmetry-breaking effects on displacement energies and charge radius
differences in mirror nuclei}

\author[inst1]{Myeong-Hwan Mun}
\author[inst2]{Kyoungsu Heo}
\author[inst2]{Jubin Park\corref{cor1}}
\ead{honolov@ssu.ac.kr}
\author[inst2]{Myung-Ki Cheoun\corref{cor1}}
\ead{cheoun@ssu.ac.kr}
\author[inst3,inst4,inst5]{H. Sagawa}
\cortext[cor1]{Corresponding authors.}

\affiliation[inst1]{
  organization={Department of Physics, Kyungpook National University},
  city={Daegu},
  postcode={41566},
  country={Republic of Korea}
}

\affiliation[inst2]{
  organization={Department of Physics and Origin of Matter and Evolution of Galaxies Institute, Soongsil University},
  city={Seoul},
  postcode={06978},
  country={Republic of Korea}
}

\affiliation[inst3]{
  organization={RIKEN, Nishina Center for Accelerator-Based Science},
  city={Wako},
  postcode={351-0198},
  country={Japan}
}

\affiliation[inst4]{
  organization={Center for Mathematics and Physics, University of Aizu},
  city={Aizu-Wakamatsu, Fukushima},
  postcode={965-8560},
  country={Japan}
}

\affiliation[inst5]{
  organization={Institute of Theoretical Physics, Chinese Academy of Sciences},
  city={Beijing},
  postcode={100190},
  country={China}
}

\begin{abstract}
We study the effect of charge symmetry breaking (CSB) energy density function (EDF) on the mirror displacement energies (MDEs) and charge radius differences of mirror nuclei within the self-consistent Hartree-Fock-Bogolyubov (HFB) model taking Skyrme EDFs, SLy4 and SkM* as the central part of nuclear potential. We introduce the volume and the derivative terms in CSB EDF and calibrate the strength adopting four reference mirror pairs $^{34}$Ar--$^{34}$S, $^{36}$Ca--$^{36}$S, $^{38}$Ca--$^{38}$Ar, and $^{54}$Ni--$^{54}$Fe, for which experimental data of both MDEs and mirror charge-radius differences are available.   We introduce a sensitivity matrix which connects two CSB terms to  residuals of MDEs and mirror charge-radius differences after subtracting  the effect of Coulomb interaction.  By using the sensitivity matrix, we found out that the derivative term is important to reproduce both observables in a good accuracy together with the volume term, especially for the residual of mirror charge radii.  The optimized CSB EDF are further applied to predict the charge radius differences of mirror pairs,  $^{40}$Ti--$^{40}$Ar, $^{42}$Ti--$^{42}$Ca, $^{46}$Cr--$^{46}$Ti, and $^{50}$Fe--$^{50}$Cr.  We pointed out also that the CSB effects change  neutron skins of mirror  proton-rich nuclei at the $10^{-2}$ fm level  so that the  CSB contributions must be included before charge-radius differences between mirror nuclei are used to extract  neutron-skin or symmetry-energy parameters.
\end{abstract}


\end{frontmatter}

\section{Introduction}
\label{sec:introduction}

Mirror nuclei provide a sensitive test of isospin symmetry in finite nuclei. In the charge-symmetric limit, the strong interaction is invariant under proton--neutron interchange, and the two members of a mirror pair have the same nuclear structure after that interchange.  Deviations from this limit arise from the Coulomb interaction, nuclear charge-symmetry breaking (CSB), charge-independence breaking, subleading electromagnetic effects, and the self-consistent polarization of the nuclear density.

The traditional energy observable {to disentangle the isospin symmetry breaking (ISB) effect} is the mirror displacement energy (MDE). Coulomb effects dominate the MDE but do not exhaust its systematics, as summarized by the Okamoto--Nolen--Schiffer anomaly~\cite{Okamoto1964,NolenSchiffer1969,Brown2000}. Recent studies have shown that class-III CSB terms account for a substantial part of the MDE difference that remains after subtracting the Coulomb contributions, whereas the class-II term, i.e., the charge independence breaking (CIB) term,  is more directly connected with triplet displacement energies~\cite{Baczyk2018,Naito2022CSBEDF,Sagawa2024}. This remaining MDE difference is not a model-independent measure of the bare CSB interaction.  It is an experiment--theory difference defined with respect to a chosen many-body framework and may absorb omitted electromagnetic, surface, shell, pairing, and deformation effects.

A second observable, which has become increasingly important in recent experiments, is the charge-radius difference between mirror nuclei $\Delta R_{\rm ch}^{\rm mirr}$.  {In the context of  the  charge-symmetry invariance, the neutron rms radius of one mirror partner is equivalent to the proton rms radius of the other}, making $\Delta R_{\rm ch}^{\rm mirr}$ a potential proxy for neutron-skin thickness and a probe of the isovector sector of the nuclear equation of state (EoS)~\cite{Brown2017,Brown2020,Pineda2021,ReinhardNazarewicz2022}. In reality,  this relation is modified by Coulomb interaction and by nuclear isospin-symmetry-breaking interactions. Quantitative energy density function (EDF) and ab initio studies have shown that these corrections can affect neutron-skin and symmetry-energy extractions, and that the correlation between mirror charge radii and the symmetry-energy slope $L$ is pair- and shell-dependent~\cite{Naito2022Radius,Novario2023,Hu2024}.

\begin{figure*}[t] 
\centering
\includeorplaceholder{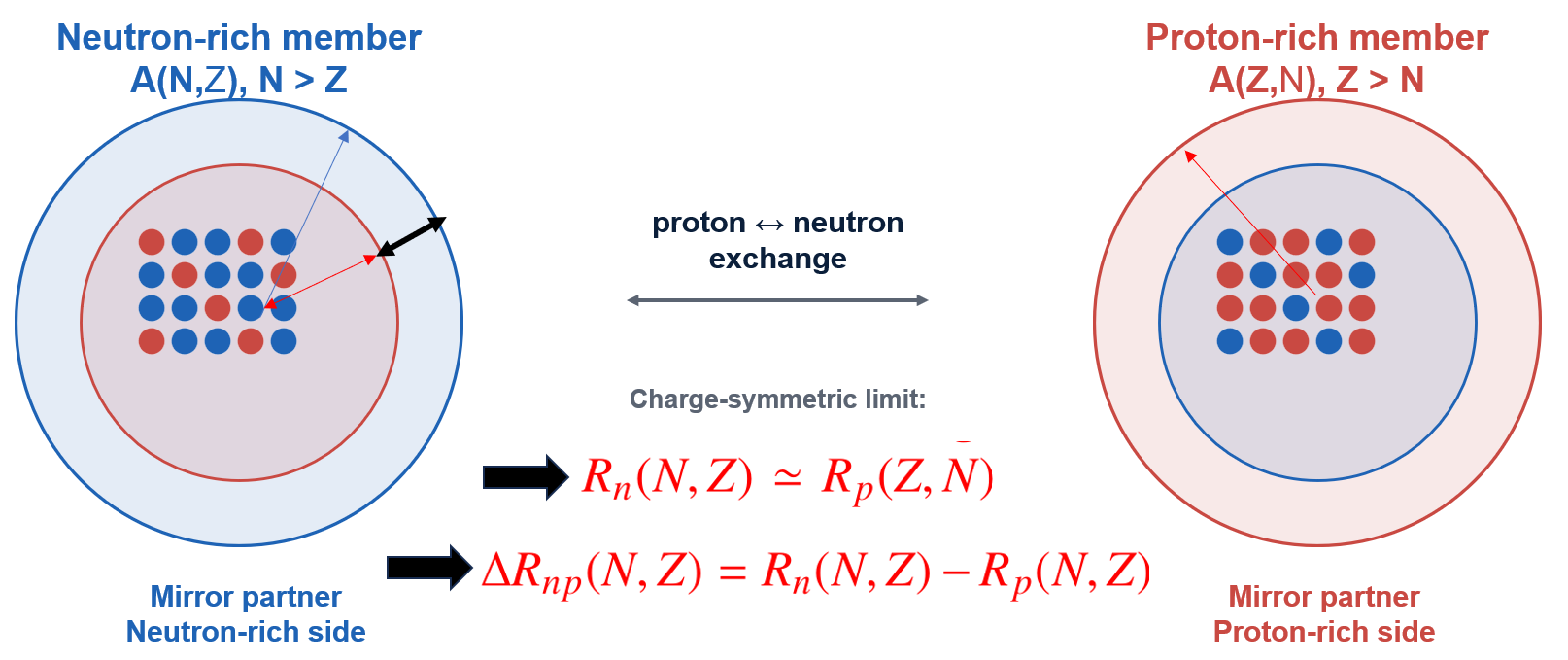}{0.92\textwidth}
\caption{{Schematic connection between mirror charge radii and neutron skin thickness.  For a neutron-rich nucleus $A(N,Z)$, the neutron skin is $\Delta R_{np}(N,Z)=R_n(N,Z)-R_p(N,Z)$.  In the charge-symmetric limit of the strong interaction, proton--neutron exchange maps this nucleus onto the proton-rich mirror partner $A(Z,N)$, giving the point-radius relation $R_n(N,Z)\simeq R_p(Z,N)$.  Thus the proton radius, and after finite-size corrections the charge radius, of the proton-rich mirror partner can provide information on the neutron skin of the neutron-rich member.  In realistic nuclei this relation is modified by Coulomb polarization and nuclear CSB/CIB effects, motivating the finite-density CSB calibration used in this work.  The drawing is schematic and not to scale.}}
\label{fig:mirror_skin_schematic}
\end{figure*}

{Figure~\ref{fig:mirror_skin_schematic} illustrates the basic mirror-radius argument used in this work. In the charge-symmetric limit, proton--neutron exchange gives $R_n(N,Z)\simeq R_p(Z,N)$, so the neutron skin of the neutron-rich member can be related, at the point-radius level, to the proton radius of the proton-rich mirror partner.}

The purpose of the present work is predictive. We first use four measured mirror pairs as a calibration set for finite-density class-III CSB interactions, which have both the volume and gradient terms. The calibration is constrained by MDEs and by measured mirror charge-radius differences within the same self-consistent EDF framework.  We then apply the calibrated CSB interactions to proton-rich mirror partners whose charge radii have not yet been measured. 

This work addresses three questions. 
First, can the four measured mirror pairs constrain a common finite-density class-III CSB correction? 
Second, does the calibrated two-coupling CSB correction simultaneously reduce MDE and mirror charge-radius residuals without introducing radius-specific parameters? 
Third, what charge radii are predicted for the unmeasured proton-rich mirror partners \(^{40}\)Ti, \(^{42}\)Ti, \(^{46}\)Cr, and \(^{50}\)Fe?

The present work differs from previous mirror-radius correlation studies in one important respect.  We do not use mirror charge-radius differences only as empirical proxies for neutron skins, nor do we use MDEs only as independent constraints on an isospin-breaking interaction.  Instead, we impose the sequence: calibration of finite-density CSB effects with MDEs, validation of their spatial content with measured charge radii, and prediction of unmeasured proton-rich charge radii. To this end, we introduce a local MDE--radius sensitivity matrix that connects the two CSB couplings to the MDE and mirror charge-radius residuals.

\section{Calibration of finite-density CSB effects}
\label{sec:calibration}

\subsection{Calibration observables, EDF reference, and CSB functional}
\label{subsec:framework}

For each mirror pair, we define the MDE and mirror charge-radius difference as
\begin{align}
  \Delta B_i &= B_i(Z,N)-B_i(N,Z),
  \label{eq:mde_def}\\
  \Delta R_{{\rm ch},i}^{\rm mirr}
  &=R_{{\rm ch},i}(Z,N)-R_{{\rm ch},i}(N,Z).
  \label{eq:radius_def}
\end{align}
Here $B$ denotes the positive binding energy, $(Z,N)$ is the proton-rich member, and $(N,Z)$ is the neutron-rich mirror partner.  A positive $\Delta R_{\rm ch}^{\rm mirr}$ means that the proton-rich partner has the larger charge radius.  The measured calibration set consists of $^{34}$Ar--$^{34}$S, $^{36}$Ca--$^{36}$S, $^{38}$Ca--$^{38}$Ar, and $^{54}$Ni--$^{54}$Fe, for which both MDEs and mirror charge-radius differences are available~\cite{Wang2021AME,AngeliMarinova2013,Miller2019,Pineda2021}.

Calculations employ a modified HFBTHO v4.0 solver~\cite{Stoitsov2005HFBTHO,Stoitsov2013HFBTHO,NavarroPerez2017HFBTHO,Marevic2022HFBTHO} with SLy4~\cite{Chabanat1998SLy4} and SkM*~\cite{Bartel1982SkM}.  The pairing prescription, basis, Coulomb treatment, initial deformations, and convergence criteria are identical for both members of each pair and for all coupling-switching calculations.  The four-pair calibration calculations use $N_{\rm sh}=16$ and an initial deformation scan $-0.20\le\beta_2^{\rm init}\le0.20$.  For the target-pair calculations, the scan is extended to $-0.30\le\beta_2^{\rm init}\le0.30$, and $N_{\rm sh}=18$ is used for the $A=42$, 46, and 50 pairs.  Further numerical details and finite-difference tests are given in the Supplementary Material.

{The charge radii are evaluated from
\begin{equation}
 R_{\rm ch}^2=R_p^2+\langle r_p^2\rangle
 +\frac{N}{Z}\langle r_n^2\rangle+R_{\rm DF}^2+R_{\rm so}^2,
 \label{eq:rch}
\end{equation}
where $R_p$ is the point-proton rms radius.  We use $\langle r_p^2\rangle=0.769~{\rm fm}^2$, $\langle r_n^2\rangle=-0.116~{\rm fm}^2$, and the Darwin-Foldy
term, $R_{\rm DF}^2=0.033~{\rm fm}^2$.  The present implementation sets $R_{\rm so}^2=0$; its expected contribution and other residual radius-model effects at this scale are represented by the adopted $0.005$ fm radius-error floor.}

The CSB-free reference calculation is
\begin{equation}
E^{(C)}=E_{\rm CI}+E_{\rm Coul}+E_{\rm pair},
\label{eq:coul_ref}
\end{equation}
{where $E_{\rm CI}$ is the charge-independent Skyrme EDF.  The Coulomb energy includes the  Hartree contribution self-consistently and the exchange contribution in the Slater approximation.} {We use the superscript $(C)$ to denote this Coulomb-included, nuclear-CSB-free reference.}  The residuals to be described by the CSB functional are
\begin{align}
 R_{B,i}&=\Delta B_i^{\rm exp}-\Delta B_i^{(C)},
 \label{eq:RB}\\
 R_{R,i}^{(C)}&=\Delta R_{{\rm ch},i}^{{\rm mirr},{\rm exp}}
 -\Delta R_{{\rm ch},i}^{{\rm mirr},(C)}.
 \label{eq:RR}
\end{align}
Here and below, a residual is defined as  a difference between  experiment and theory.  For both EDFs, the reference calculation yields $R_{B,i}<0$ and $R_{R,i}^{(C)}<0$ for all four calibration pairs, indicating that it underestimates the magnitude of the MDE and overestimates $\Delta R_{\rm ch}^{\rm mirr}$.

After fixing the reference calculation, we add an explicit class-III CSB functional consisting of a volume term and a surface-gradient term,
\begin{align}
 {\cal H}_{\rm CSB}^{(0)}
 &=\frac{1}{2}t_0^{\rm III}\rho_0\rho_1
 =\frac{1}{2}t_0^{\rm III}(\rho_n^2-\rho_p^2),
 \label{eq:csb_volume}\\
 {\cal H}_{\rm CSB}^{(\Delta)}
 &=C_\Delta^{\rm III}
 \left(\rho_0\Delta\rho_1+\rho_1\Delta\rho_0\right),
 \label{eq:csb_surface}
\end{align}
with $\rho_0=\rho_n+\rho_p$ and $\rho_1=\rho_n-\rho_p$.  The volume term corresponds to the leading-order zero-range class-III CSB interaction used in Skyrme-EDF analyses of MDEs~\cite{Baczyk2018,Suzuki1993_PRC_CSB_CIB}, while the surface-gradient term is used as a reduced EDF representation of the next-to-leading-order class-III Skyrme ISB gradient terms~\cite{Baczyk2019_JPG_IMME_NLO_ISB}.  Since
\begin{equation}
\rho_0\Delta\rho_1+\rho_1\Delta\rho_0
=2(\rho_n\Delta\rho_n-\rho_p\Delta\rho_p),
\end{equation}
integration by parts gives a contribution proportional to $-2[(\nabla\rho_n)^2-(\nabla\rho_p)^2]$. The term in Eq.~(\ref{eq:csb_volume}) probes the spatially integrated product $\rho_0\rho_1$, i.e., the bulk-weighted neutron-proton density imbalance, whereas the term in Eq.~(\ref{eq:csb_surface}) is sensitive to  changes of  radial shape in which the CSB field acts. This distinction is essential because MDEs are integrated energy observables while mirror radii and neutron skins are surface sensitive.

This volume-plus-surface-gradient CSB structure also has a covariant analogue. In the RMF approach of Ref.~\cite{Tanimura2025}, class-III nuclear CSB is generated by $\omega-\rho^0$ meson mixing.  Eliminating the finite-range meson fields, or equivalently expanding the meson propagators in gradients, produces leading structures proportional to $\rho_V\rho_{TV}$ and $\rho_V\Delta\rho_{TV}+\rho_{TV}\Delta\rho_V$.  These are the covariant counterparts of the $\rho_0\rho_1$ and $\rho_0\Delta\rho_1+\rho_1\Delta\rho_0$ terms used here.

\subsection{Local sensitivity matrix and effective CSB combinations}
\label{subsec:response}

For each EDF and each measured mirror pair, the local response to the two CSB couplings is described by the sensitivity matrix
\begin{equation}
  {\bf S}_i=
  \begin{pmatrix}
  S_{0,i}^{B} & S_{\Delta,i}^{B}\\
  S_{0,i}^{R} & S_{\Delta,i}^{R}
  \end{pmatrix}.
  \label{eq:S_matrix}
\end{equation}
It is a local Jacobian evaluated at the CSB-free reference point. as follows.   In the linear approximation,
\begin{equation}
 \begin{pmatrix}
  \delta_{\rm CSB}\Delta B_i\\[2pt]
  \delta_{\rm CSB}\Delta R_{{\rm ch},i}^{\rm mirr}
 \end{pmatrix}
 \simeq
 {\bf S}_i
 \begin{pmatrix}
  t_0^{\rm III}\\[2pt]
  C_\Delta^{\rm III}
 \end{pmatrix}.
 \label{eq:matrix_response}
\end{equation}
Here $\delta_{CSB} {\rm O}$ for any quantity ${\rm O}$ is defined as $\delta_{CSB} {\rm O} \equiv \rm{O(C+CSB) - O(C)}$. Thus $S_{0,i}^{B}$ and $S_{\Delta,i}^{B}$ measure the MDE sensitivities to the volume and surface-gradient CSB couplings, while $S_{0,i}^{R}$ and $S_{\Delta,i}^{R}$ measure the corresponding sensitivities of the mirror charge-radius difference. These four coefficients are calculated from self-consistent HFBTHO solutions without introducing any adjustable parameters. 

{The MDE part of Eq.~\eqref{eq:matrix_response} can be written as
\begin{equation}
\delta_{\rm CSB} \Delta B_{{\rm CSB},i}
\simeq
S_{0,i}^{B}t_0^{\rm III}+S_{\Delta,i}^{B}C_\Delta^{\rm III}
=
S_{0,i}^{B}\left(t_0^{\rm III}-\lambda_i^B C_\Delta^{\rm III}\right),
\label{eq:mde_response}
\end{equation}
with
\begin{equation}
\lambda_i^B=-\frac{S_{\Delta,i}^{B}}{S_{0,i}^{B}}.
\label{eq:lambdaB_def}
\end{equation}
Similarly, the mirror-radius part is
\begin{equation}
\delta_{\rm CSB}\Delta R_{{\rm ch},i}^{\rm mirr}
\simeq
S_{0,i}^{R}t_0^{\rm III}+S_{\Delta,i}^{R}C_\Delta^{\rm III}
=
S_{0,i}^{R}\left(t_0^{\rm III}-\lambda_i^R C_\Delta^{\rm III}\right),
\label{eq:radius_response}
\end{equation}
with
\begin{equation}
\lambda_i^R=-\frac{S_{\Delta,i}^{R}}{S_{0,i}^{R}}.
\label{eq:lambdaR_def}
\end{equation}

The quantities $\lambda_i^B$ and $\lambda_i^R$ are therefore diagnostic ratios of the coupling sensitivities to MDE and charge radius difference: they lead the two columns of ${\bf S}_i$ into a direction of a single column in the $(t_0^{\rm III},C_\Delta^{\rm III})$ coupling plane.  This process may elucidate  a relative role of the volume and derivative terms quantitatively and find out whether the two observables are enough to calibrate magnitudes of the two terms 
from the experimental observables.

{The local slopes are obtained by central finite differences.  For a general observable $O_i$,
\begin{align}
S_{0,i}^{O} &= \frac{O_i(+h_0,0)-O_i(-h_0,0)}{2h_0}, \nonumber\\
S_{\Delta,i}^{O} &= \frac{O_i(0,+h_\Delta)-O_i(0,-h_\Delta)}{2h_\Delta}.
\label{eq:finite_difference_general}
\end{align}
The steps are $h_0=5.6~{\rm MeV\,fm^3}$ and $h_\Delta=1.0~{\rm MeV\,fm^5}$.  For each $+h$ and $-h$ value, both mirror members are solved again to self-consistency with the same coupling value; the mirror difference is then formed from the two resulting nuclei. }

{The complete set of local sensitivity coefficients used to construct
\(\lambda_i^B\) and \(\lambda_i^R\) is given in Supplementary Table~S4.
That table lists \(S_{0,i}^{B}\), \(S_{\Delta,i}^{B}\),
\(S_{0,i}^{R}\), and \(S_{\Delta,i}^{R}\) for each measured calibration
pair and for both EDFs. It should be read as the numerical form of the
local Jacobian in Eq.~(10); the full fit uses these full sensitivity
coefficients, not the compressed ratios alone.
The comparison of $\lambda_i^B$ and $\lambda_i^R$ is a critical reference to clarify the role of the surface-gradient term.  If the two ratios were identical for all calibration pairs, MDEs and mirror radii would probe the same effective CSB combination which suggests no possibility of determining the surface-gradient CSB term independently.  In other words, a difference between the two ratios shows that mirror charge radii test a different radial component of the CSB field.  

For example, in the SLy4 calculation for $^{34}$Ar--$^{34}$S,
\begin{align}
S_{0}^{B}&=+0.14575~{\rm fm}^{-3}, & \nonumber\\
S_{\Delta}^{B}&=-0.18602~{\rm fm}^{-5}, & \nonumber\\
S_{0}^{R}&=-2.7777\times10^{-4}~{\rm MeV}^{-1}{\rm fm}^{-2}, & \nonumber \\
S_{\Delta}^{R}&=+6.5450\times10^{-4}~{\rm MeV}^{-1}{\rm fm}^{-4},
\label{eq:sl4_34_example_S}
\end{align}
which gives
\begin{equation}
\lambda^B=1.276,
\qquad
\lambda^R=2.356.
\label{eq:sl4_34_example_lambda}
\end{equation}
The MDE and mirror-radius directions are therefore not parallel in this representative case.  This observation explains why the surface-gradient term is not introduced merely to reparametrize the MDE, but to allow simultaneous control of the energy and radius observables.  The full four-pair comparison should be made with both $\lambda_i^B$ and $\lambda_i^R$. The representative \(^{34}\mathrm{Ar}\)--\(^{34}\mathrm{S}\) example
illustrates the interpretation of one row of the full sensitivity table.
The complete four-pair comparison is reported in Supplementary Table~S4
and summarized in Table~2 through the two diagnostic ratios
\(\lambda_i^B\) and \(\lambda_i^R\). The comparison of \(\lambda_i^B\) and \(\lambda_i^R\) is more than a check of the stability of \(t_{\rm eff}^{\rm III}\). 
It directly tests whether MDEs and mirror charge radii require different combinations of the volume and surface-gradient CSB terms.

When the MDE ratios $\lambda_i^B$ are close for the calibration pairs, the MDEs are sensitive mainly to the effective combination
\begin{equation}
 t_{\rm eff}^{\rm III}=t_0^{\rm III}-\lambda_{\rm cl}^{B}C_\Delta^{\rm III},
 \label{eq:teff}
\end{equation}
rather than to $t_0^{\rm III}$ and $C_\Delta^{\rm III}$ separately.  The corresponding complementary diagnostic coordinate,
\begin{equation}
 t_{\perp}^{\rm III}=t_0^{\rm III}+\lambda_{\rm cl}^{B}C_\Delta^{\rm III},
 \label{eq:tperp}
\end{equation}
only describes motion along the weakly constrained covariance valley and should not be interpreted as an independent physical coupling.  {The existence of this MDE-compatible covariance band does not imply that the surface-gradient term is redundant. 
Rather, it shows that MDEs alone cannot determine the radial structure of the CSB field; the comparison with \(\lambda_i^R\) supplies the constraint  on the radius dependence of the sensitivity matrix.

The fitted parameters minimize
\begin{equation}
\chi^2=
\sum_{i\in{\cal C}}\left(\frac{R_{B,i}^{(C+{\rm CSB})}}{\sigma_{B,i}}\right)^2
+
\eta_R
\sum_{i\in{\cal C}}\left(\frac{R_{R,i}^{(C+{\rm CSB})}}{\sigma_{R,i}}\right)^2,
\label{eq:chi2}
\end{equation}
where ${\cal C}$ denotes the four measured calibration pairs.  The uncertainties are
\begin{align}
\sigma_{B,i}^2 &= (\sigma_{B,i}^{\rm exp})^2 + (0.100~{\rm MeV})^2, \nonumber\\
\sigma_{R,i}^2 &= (\sigma_{R,i}^{\rm exp})^2 + (0.005~{\rm fm})^2.
\label{eq:sigmas}
\end{align}
We use $\eta_R=0$ for MDE-only fits and $\eta_R=1$ for full mass-radius fits. 
 No extra parameter is introduced in the radius sector.  The full fit determines a common coupling point from the full set of MDE and 
 radius sensitivities; it is not a fit to $\lambda_i^B$ or $\lambda_i^R$ alone. {Table~1 summarizes the three calibration protocols 
 considered here: a volume-only MDE fit, a volume-plus-surface-gradient MDE fit, and a full MDE-radius fit.}

\begin{table*}[t] 
\centering
\caption{Calibration performance for the four measured mirror pairs.  The table shows how volume-only, volume-plus-surface-gradient, and full mass-radius calibrations determine the class-III CSB couplings.  The effective combination turns out to be $t_{\rm eff}^{\rm III}=t_0^{\rm III}-\lambda_{\rm cl}C_\Delta^{\rm III}$.  The quantities RMS$_B$ and RMS$_R$ are unweighted RMS residuals of the MDEs and mirror charge-radius differences, respectively, over the four calibration pairs.}
\label{tab:fit_summary_anchor}
\setlength{\tabcolsep}{5pt}
\begin{tabular}{llcccccc}
\hline
EDF
& fit
& $t_0^{\rm III}$
& $C_\Delta^{\rm III}$
& $t_{\rm eff}^{\rm III}$
& $\chi^2/{\rm dof}$
& RMS$_B$
& RMS$_R$
\\
&
& [MeV fm$^{3}$]
& [MeV fm$^{5}$]
& [MeV fm$^{3}$]
&
& [MeV]
& [fm]
\\
\hline
SLy4
& $t_0$, mass
& $-6.51(26)$
& --
& $-6.51$
& 0.504
& 0.0625
& 0.0083
\\
& $t_0+C_\Delta$, mass
& $-17.40(11.10)$
& $-8.30(8.40)$
& $-6.47$
& 0.271
& 0.0370
& 0.0052
\\
& $t_0+C_\Delta$, full
& $-25.80(7.90)$
& $-14.70(6.00)$
& $-6.42$
& 0.368
& 0.0529
& 0.0031
\\
\hline
SkM*
& $t_0$, mass
& $-6.67(27)$
& --
& $-6.67$
& 0.290
& 0.0479
& 0.0185
\\
& $t_0+C_\Delta$, mass
& $-10.81(6.33)$
& $-3.49(5.33)$
& $-6.62$
& 0.222
& 0.0334
& 0.0174
\\
& $t_0+C_\Delta$, full
& $-24.73(5.57)$
& $-15.25(4.69)$
& $-6.42$
& 3.919
& 0.1172
& 0.0136
\\
\hline
\end{tabular}
\end{table*}

The four measured calibration pairs define a compact $\lambda_B$ band in both EDFs, corresponding to a surface-gradient-dominated finite-density CSB pattern: $\lambda_{\rm cl}^{\rm SLy4}=1.32\pm0.04~{\rm fm}^{-2}$ and $\lambda_{\rm cl}^{\rm SkM^*}=1.20\pm0.05~{\rm fm}^{-2}$.  The quoted uncertainties are pair-to-pair sample standard deviations, not statistical fit errors.  The compact $\lambda_B$ band is recovered in both EDFs, although the mean value changes by about 10\%.

The pair-by-pair validation is shown in Table~\ref{tab:calibration_validation}.  The radius residual after adding CSB is
\begin{equation}
 R_{R,i}^{(C+{\rm CSB})}=\Delta R_{{\rm ch},i}^{{\rm mirr},{\rm exp}}
 -\Delta R_{{\rm ch},i}^{{\rm mirr},(C+{\rm CSB})}.
 \label{eq:RRcsb}
\end{equation}
A reduction of $|R_{R,i}|$ after an MDE-compatible CSB correction indicates that the CSB correction has the appropriate radial structure.  In SLy4, the full fit reduces the radius RMS to 0.0031 fm and shifts the signed neutron-proton radius difference of the proton-rich partners by $0.011$--$0.019$ fm.  SkM* retains a compatible $\lambda_B$ band but leaves a larger radius RMS, 0.0136 fm, and gives shifts of $0.015$--$0.026$ fm.  This comparison shows that the effective CSB direction is more robust than the absolute radius correction, which depends on the reference EDF.

\begin{revision}
\begin{table*}[t] 
\centering
\caption{
{Pair-by-pair calibration diagnostics for the four measured mirror pairs.
\(R_B\) is the CSB-free MDE residual, and \(R_R^{(C)}\) and
\(R_R^{(C+{\rm CSB})}\) are mirror charge-radius residuals before and after
the full fit. The diagnostic ratios \(\lambda_i^B\) and \(\lambda_i^R\)
are obtained from the local sensitivity coefficients tabulated in
Supplementary Table~S4:
\(\lambda_i^B=-S_{\Delta,i}^{B}/S_{0,i}^{B}\) for the MDE and
\(\lambda_i^R=-S_{\Delta,i}^{R}/S_{0,i}^{R}\) for the mirror charge-radius
difference. Their comparison tests whether the energy and radius observables
probe the same or different effective coupling combinations in the
\((t_0^{\rm III},C_\Delta^{\rm III})\) coupling plane. The final column
denotes the CSB-induced change in the neutron--proton radius
difference \(\Delta R_{np}^{p}=R_n^{p}-R_p^{p}\) of the proton-rich partner.
}}
\label{tab:calibration_validation}
\setlength{\tabcolsep}{5pt}
\begin{tabular}{llcccccc}
\toprule
EDF & Pair & $R_B$ & $\lambda_B$ & {$\lambda_R$} & $R_R^{(C)}$ & $R_R^{(C+\mathrm{CSB})}$ & $\delta_{\rm CSB}\Delta R_{np}^{p}$ \\
 & & [MeV] & [fm$^{-2}$] & [fm$^{-2}$] & [fm] & [fm] & [fm] \\
\midrule
SLy4 & $^{34}$Ar--$^{34}$S & $-0.962$ & 1.276 & 2.356 & $-0.0037$ & $-0.0012$ & $+0.0189$ \\
 & $^{36}$Ca--$^{36}$S & $-2.009$ & 1.295 & 2.377 & $-0.0075$ & $-0.0022$ & $+0.0179$ \\
 & $^{38}$Ca--$^{38}$Ar & $-1.008$ & 1.344 & 2.393 & $-0.0084$ & $-0.0057$ & $+0.0163$ \\
 & $^{54}$Ni--$^{54}$Fe & $-1.032$ & 1.359 & 2.180 & $-0.0016$ & $-0.0002$ & $+0.0111$ \\
\midrule
SkM* & $^{34}$Ar--$^{34}$S & $-0.887$ & 1.186 & 2.049 & $-0.0103$ & $-0.0085$ & $+0.0259$ \\
 & $^{36}$Ca--$^{36}$S & $-1.970$ & 1.146 & 2.121 & $-0.0242$ & $-0.0196$ & $+0.0241$ \\
 & $^{38}$Ca--$^{38}$Ar & $-0.994$ & 1.194 & 2.004 & $-0.0179$ & $-0.0162$ & $+0.0216$ \\
 & $^{54}$Ni--$^{54}$Fe & $-1.109$ & 1.275 & 1.860 & $-0.0048$ & $-0.0039$ & $+0.0154$ \\
\bottomrule
\end{tabular}
\end{table*}
\end{revision}

\begin{figure*}[t] 
\centering
\includeorplaceholder{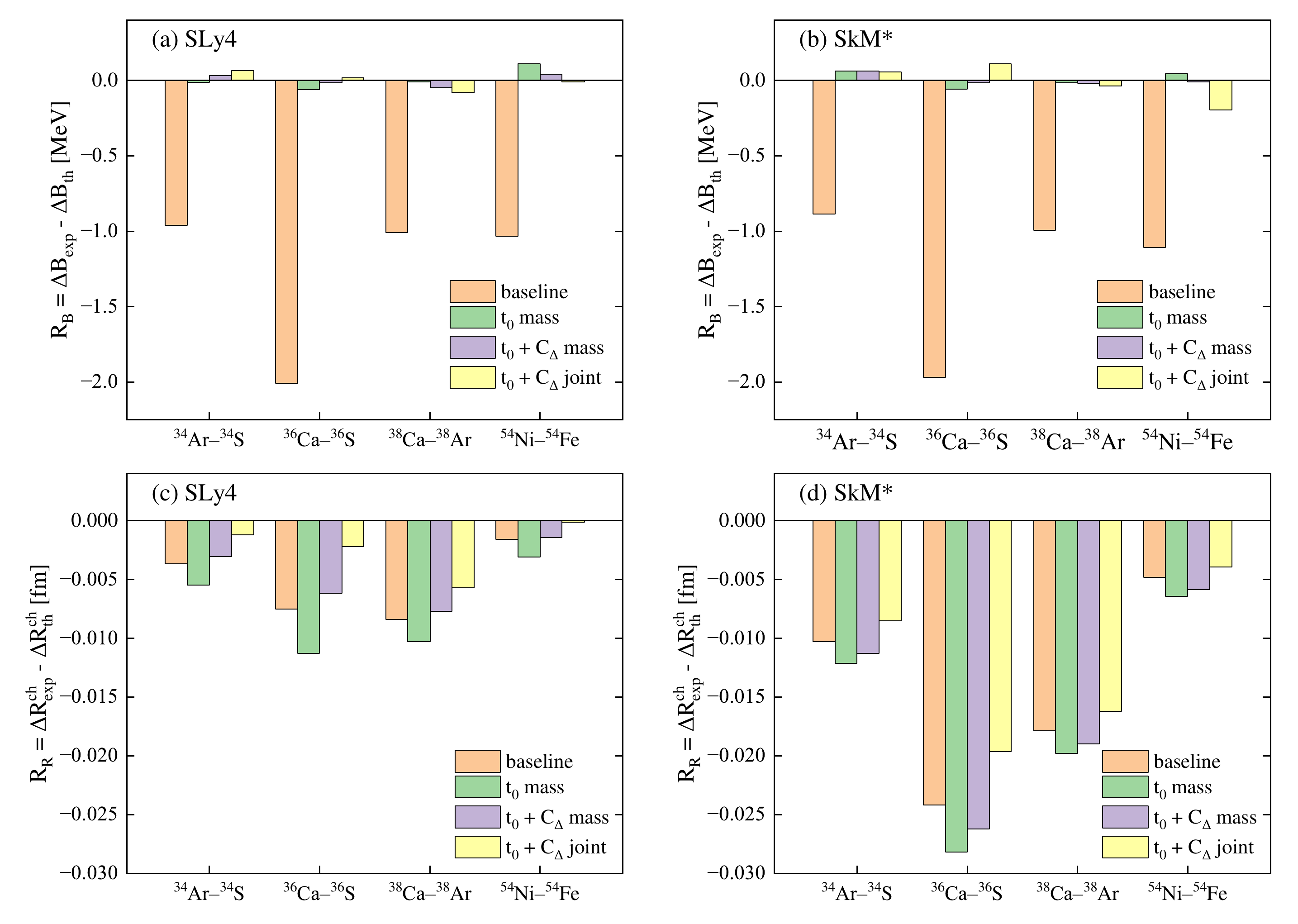}{0.95\textwidth}
\caption{MDE and mirror charge-radius residuals for the four measured calibration pairs. Panels (a) and (b) show the MDE residuals,
\(R_{B,i}=\Delta B_i^{\rm exp}-\Delta B_i^{\rm th}\), for SLy4 and SkM*, respectively. Panels (c) and (d) show the corresponding mirror charge-radius residuals,
\(R_{R,i}=\Delta R_{{\rm ch},i}^{\rm mirr,exp} -\Delta R_{{\rm ch},i}^{\rm mirr,th}\).
For each pair, the bars denote the CSB-free reference calculation, the volume-only MDE calibration, the volume-plus-surface-gradient MDE calibration, and the full mass--radius calibration. Values closer to zero indicate better agreement with experiment.
The upper panels show how the CSB terms reduce the energy residuals, while the lower panels test whether the same CSB correction gives the required spatial change in mirror charge radii.
}
\label{fig:residuals}
\end{figure*}

Figure~\ref{fig:residuals} visualizes the same calibration diagnostics in residual form.
The upper panels show the MDE residuals for SLy4 and SkM*, while the lower panels show the corresponding mirror charge-radius residuals.
For each pair, the bars compare the CSB-free reference calculation, the volume-only MDE calibration, the volume-plus-surface-gradient MDE calibration, and the full mass--radius calibration.
The MDE panels show how the CSB terms reduce the remaining energy differences, whereas the radius panels provide an independent spatial test of the same CSB correction. In SLy4, the full calibration moves the charge-radius residuals closer to zero for all four calibration pairs.
In SkM*, the MDE residuals are also reduced, but the radius residuals remain larger, indicating stronger EDF dependence in the propagation from the calibrated CSB couplings to charge radii.

In brief, the SkM* MDE RMS, radius RMS, and the covariance between $t_0^{\rm III}$ and $C_\Delta^{\rm III}$ changed only weakly, indicating that the SkM* mass-radius tension is not removed by a small missing-electromagnetic offset. Because only two reference EDFs are considered, their difference is used as a limited diagnostic of model dependence, not as a statistically complete EDF uncertainty.

\section{Predictions for unmeasured proton-rich charge radii}
\label{sec:predictions}

{We now use the calibrated finite-density CSB effects to predict charge radii} for proton-rich mirror partners not used in the calibration.  For a target pair $j$, the predicted mirror charge-radius difference and the proton-rich radius are
\begin{align}
 \Delta R_{{\rm ch},j}^{{\rm mirr},{\rm pred}}
 &=\Delta R_{{\rm ch},j}^{{\rm mirr},(C)}
 +\delta_{\rm CSB}\Delta R_{{\rm ch},j}^{\rm mirr},
 \label{eq:pred_difference}\\
 R_{{\rm ch},j}^{p,{\rm pred}}
 &=R_{{\rm ch},j}^{n,{\rm exp}}
 +\Delta R_{{\rm ch},j}^{{\rm mirr},{\rm pred}}.
 \label{eq:pred_radius}
\end{align}
Here $p$ and $n$ denote the proton-rich and neutron-rich mirror partners, respectively.  Neither the target-pair MDEs nor the proton-rich target radii enter the calibration.  The experimental radii $R_{{\rm ch},j}^{n,{\rm exp}}$ are taken from Ref.~\cite{AngeliMarinova2013}.

The target predictions are restricted to pairs whose calculated $\lambda_B$ values remain compatible with the compact calibration-pair $\lambda_B$ band.  The four selected targets, $^{40}$Ti--$^{40}$Ar, $^{42}$Ti--$^{42}$Ca, $^{46}$Cr--$^{46}$Ti, and $^{50}$Fe--$^{50}$Cr, satisfy this criterion in both EDFs.  Their mean $\lambda_B$ values are $1.22\pm0.03~{\rm fm}^{-2}$ in SLy4 and $1.17\pm0.01~{\rm fm}^{-2}$ in SkM*.  Additional low-$\lambda_B$ mirror pairs are used only as diagnostic checks in the Supplementary Material and are not included in the target set.

The uncertainties in parentheses in Table~\ref{tab:predictions} are the minimal experimental-plus-radius-floor uncertainties,
\begin{equation}
    \sigma_{{\rm min},j}
    =
    \left[
        \left(\sigma_{{\rm exp},j}^{n}\right)^2
        +
        (0.005~{\rm fm})^2
    \right]^{1/2}.
    \label{eq:sigma_min}
\end{equation}
We quote separately
\begin{equation}
    \delta_{{\rm EDF},j}
    =
    \left|
        \Delta R_{{\rm ch},j}^{\rm SLy4}
        -
        \Delta R_{{\rm ch},j}^{\rm SkM^*}
    \right|
    \label{eq:delta_edf}
\end{equation}
as a two-EDF difference and a limited diagnostic of model dependence, not as a statistical standard deviation.  Coupling-covariance and $\lambda_B$-band assignment uncertainties are not included in the parenthetical errors.

\begin{table*}[t] 
\centering
\caption{Predicted charge radii of proton-rich mirror partners from the jointly calibrated finite-density CSB effects.  Entries in the $\lambda_B$ and $\Delta R_{\rm ch}^{\rm mirr,pred}$ columns are shown as SLy4/SkM*.  Absolute radii use the SLy4 central value.  Parentheses combine the experimental uncertainty of the known neutron-rich partner with the 0.005 fm radius-error floor.  The final column gives $\delta_{\rm EDF}=|\Delta R_{\rm ch}^{\rm SLy4}-\Delta R_{\rm ch}^{\rm SkM^*}|$, a two-EDF difference rather than a statistical standard deviation.}
\label{tab:predictions}
\setlength{\tabcolsep}{5pt}
\begin{tabular}{lccccc}
\toprule
Target pair & $\lambda_B$ & $R_{\rm ch}^{n,{\rm exp}}$ & $\Delta R_{\rm ch}^{{\rm mirr},{\rm pred}}$ & $R_{\rm ch}^{p,{\rm pred}}$ & $\delta_{\rm EDF}$ \\
 & [fm$^{-2}$] & [fm] & [fm] & [fm] & [fm] \\
\midrule
$^{40}$Ti--$^{40}$Ar & $1.248/1.173$ & $3.4274(26)$ & $0.1386/0.1466$ & $3.5660(56)$ & 0.0080 \\
$^{42}$Ti--$^{42}$Ca & $1.184/1.174$ & $3.5081(21)$ & $0.0699/0.0679$ & $3.5780(54)$ & 0.0020 \\
$^{46}$Cr--$^{46}$Ti & $1.241/1.181$ & $3.6070(22)$ & $0.0831/0.0619$ & $3.6901(55)$ & 0.0212\\
$^{50}$Fe--$^{50}$Cr & $1.198/1.168$ & $3.6588(65)$ & $0.0522/0.0545$ & $3.7110(82)$ & 0.0023 \\
\bottomrule
\end{tabular}
\end{table*}

Among the four predictions, $^{42}$Ti is the cleanest near-term test of the calibrated CSB effects because the SLy4--SkM* spread is only $0.0020$ fm.  The $^{50}$Fe prediction is similarly stable, although its mirror-radius signal is smaller.  The $^{40}$Ti pair has the largest predicted mirror-radius difference and therefore gives the strongest signal, but proton-rich structure and continuum effects may be more important.  The $^{46}$Cr prediction carries the largest EDF spread, $0.0212$ fm, and is the most useful stress test of model dependence.  These entries are targeted predictions for future charge-radius measurements, not a global charge-radius table.

\section{Summary and conclusions}
\label{sec:conclusions}
We have calibrated class-III CSB EDF corrections containing volume and surface-gradient terms using the MDEs and mirror charge-radius differences of four measured mirror pairs, and applied the calibrated corrections to predict charge radii of unmeasured proton-rich mirror partners.
The calibration pairs $^{34}$Ar--$^{34}$S, $^{36}$Ca--$^{36}$S, $^{38}$Ca--$^{38}$Ar, and $^{54}$Ni--$^{54}$Fe define a compact $\lambda_B$ band in both SLy4 and SkM*, indicating a surface-gradient-dominated CSB pattern.  Their MDEs primarily constrain the effective combination $t_{\rm eff}^{\rm III}=t_0^{\rm III}-\lambda_{\rm cl}C_\Delta^{\rm III}$, rather than a unique decomposition into independently determined volume and surface-gradient CSB couplings.

{Using the optimized CSB EDFs through the full mass-radius calibration processes,} we predict 
$R_{\rm ch}(^{40}{\rm Ti})=3.5660(56)$ fm, 
$R_{\rm ch}(^{42}{\rm Ti})=3.5780(54)$ fm, 
$R_{\rm ch}(^{46}{\rm Cr})=3.6901(55)$ fm, and 
$R_{\rm ch}(^{50}{\rm Fe})=3.7110(82)$ fm.
The $^{42}$Ti and $^{50}$Fe cases provide the most stable near-term tests, while $^{46}$Cr provides the strongest test of EDF dependence.  The predicted mirror-radius differences retain a two-EDF spread of up to 0.021 fm, which should be read as a diagnostic lower bound rather than a complete theoretical uncertainty.

The calibrated CSB corrections change proton-rich mirror skins at the $10^{-2}$ fm level.  Thus future mirror charge-radius measurements will not only test the predicted radii but also probe the finite-density CSB effects needed for neutron-skin and symmetry-energy studies.  A successful MDE calibration alone does not guarantee a unique charge-radius correction; the spatial propagation of the calibrated CSB effects must be validated in the same EDF framework.  A natural covariant extension is provided by RMF calculations with $\omega-\rho^0$ mixing, which generate covariant analogues of the volume and surface-gradient CSB structures used here.

The central implication is that finite-density isospin-breaking effects constrained by MDEs must be quantified in the same framework before $\Delta R_{\rm ch}^{\rm mirr}$ is interpreted as a clean neutron-skin or symmetry-energy probe.

\section*{Declaration of competing interest}
The authors declare that they have no known competing financial interests or personal relationships that could have appeared to influence the work reported in this paper.

\section*{Data availability}
The numerical data supporting the findings of this study are available from the corresponding authors upon reasonable request.

\section*{Acknowledgements}
M.-H.M. was supported by the National Research Foundation of Korea (NRF) grant funded by the Korean government (MSIT) under Grant Nos.~RS-2026-25487837 and RS-2018-NR031074. 
K.~Heo and M.-K.C. acknowledge support from the NRF under Grant No. RS-2024-00460031. 
M.-K.C. was also supported by the NRF Basic Science Research Program under Grant Nos. RS-2021-NR060129 and RS-2025-16071941.
J.-B.P. acknowledges support from NRF grants funded by the Korean government (MSIT) under Grant No.~RS-2025-24533596 and by the Ministry of Education under Grant No.~RS-2025-25400847. H.S. is supported by the JSPS Grant-in-Aid for Scientific Research (C) under Grant No. JP26K07079.

\end{document}


\setcounter{equation}{0}
\setcounter{table}{0}
\setcounter{figure}{0}
\renewcommand{\theequation}{S\arabic{equation}}
\renewcommand{\thetable}{S\arabic{table}}
\renewcommand{\thefigure}{S\arabic{figure}}
\begin{frontmatter}

\title{Supplementary material for
``Charge-symmetry-breaking effects on displacement energies and charge radius
differences in mirror nuclei''}

\author[inst1]{Myeong-Hwan Mun}
\author[inst2]{Kyoungsu Heo}
\author[inst2]{Jubin Park}
\author[inst2]{Myung-Ki Cheoun}
\author[inst3,inst4,inst5]{H. Sagawa}

\affiliation[inst1]{
  organization={Department of Physics, Kyungpook National University},
  city={Daegu},
  postcode={41566},
  country={Republic of Korea}
}

\affiliation[inst2]{
  organization={Department of Physics and Origin of Matter and Evolution
  of Galaxies Institute, Soongsil University},
  city={Seoul},
  postcode={06978},
  country={Republic of Korea}
}

\affiliation[inst3]{
  organization={RIKEN, Nishina Center for Accelerator-Based Science},
  city={Wako},
  postcode={351-0198},
  country={Japan}
}

\affiliation[inst4]{
  organization={Center for Mathematics and Physics, University of Aizu},
  city={Aizu-Wakamatsu, Fukushima},
  postcode={965-8560},
  country={Japan}
}

\affiliation[inst5]{
  organization={Institute of Theoretical Physics,
  Chinese Academy of Sciences},
  city={Beijing},
  postcode={100190},
  country={China}
}

\end{frontmatter}

\section{Numerical implementation}
\label{sec:supp_numerics}

The calculations use a modified HFBTHO v4.0 solver
\cite{Marevic2022HFBTHO}.  The added module evaluates the class-III
volume and $\rho\Delta\rho$ surface-gradient terms independently of the
standard Skyrme functional.  For every coupling switch and for both
members of a mirror pair, the HFB equations are solved again to
self-consistency; hence the extracted sensitivities include the induced
density polarization.  The lowest-energy converged solution among the
initial deformations listed in Table~\ref{tab:supp_setup} is retained.

\begin{table*}[t]
\centering
\caption{Numerical settings.  Settings common to the two EDFs are used
for both members of each mirror pair.}
\label{tab:supp_setup}
\small
\begin{tabular}{p{0.28\textwidth}p{0.66\textwidth}}
\toprule
Quantity & Setting \\
\midrule
Solver and EDFs
& Modified HFBTHO v4.0; SLy4 and SkM*
  \cite{Marevic2022HFBTHO,Chabanat1998SLy4,Bartel1982SkM} \\
Basis
& Axially deformed harmonic-oscillator basis, THO off;
  automatic oscillator length \\
Four-pair calibration calculations
& $N_{\rm sh}~=~16$;~~$\beta_2^{\rm init}~=~-0.20,~-0.10,~0.00,~0.10,~0.20$ \\
Target-pair calculations
& $N_{\rm sh}~=~16$ for $A~=~40$ and $N_{\rm sh}~=~18$ for
  $A~=~42,~46,~50$; \\
&  
  $\beta_2^{\rm init}=-0.30,~-0.20,~-0.10,~0.00,~0.10,~0.20,~0.30$ \\
Pairing
& Mixed density-dependent delta pairing,
  $V_n=V_p=-300~\mathrm{MeV\,fm^3}$, \\
& Mixed-pairing parameter 0.5,
  quasiparticle cutoff 60 MeV \\
Coulomb and convergence
& Direct Coulomb self-consistent, Slater exchange;
  accuracy $10^{-9}$, at most 300 iterations \\
Sensitivity steps
& $h_0=5.6~\mathrm{MeV\,fm^3}$ and
  $h_\Delta=1.0~\mathrm{MeV\,fm^5}$ \\
Fit error floors
& $0.100$ MeV for MDEs and $0.005$ fm for charge-radius differences \\
\bottomrule
\end{tabular}
\end{table*}

Charge radii are obtained from the point-proton radius using
$\langle r_p^2\rangle=0.769~\mathrm{fm^2}$,
$\langle r_n^2\rangle=-0.116~\mathrm{fm^2}$, and
$R_{\rm DF}^2=0.033~\mathrm{fm^2}$.  The present implementation sets
$R_{\rm so}^2=0$; its expected contribution and other residual
radius-model effects at this scale are represented by the adopted
$0.005$ fm radius-error floor.

For reproducibility, the Skyrme-force parameters passed to HFBTHO are
given in Table~\ref{tab:supp_skyrme}.  The spin-orbit strength is quoted
as $W_0$, corresponding to $b_4=b_4'=W_0/2$ in the HFBTHO output.

\begin{table*}[t]
\centering
\caption{Skyrme parameter sets used in the calculations.}
\label{tab:supp_skyrme}
\scriptsize
\resizebox{\textwidth}{!}{%
\begin{tabular}{lcccccccccc}
\toprule
EDF
& $t_0$ & $t_1$ & $t_2$ & $t_3$
& $x_0$ & $x_1$ & $x_2$ & $x_3$
& $\alpha$ & $W_0$ \\
\midrule
SLy4
& $-2488.913$ & $486.818$ & $-546.395$ & $13777.0$
& $0.834$ & $-0.344$ & $-1.000$ & $1.354$
& $1/6$ & $123.0$ \\
SkM*
& $-2645.000$ & $410.000$ & $-135.000$ & $15595.0$
& $0.090$ & $0.000$ & $0.000$ & $0.000$
& $1/6$ & $130.0$ \\
\bottomrule
\end{tabular}%
}
\begin{flushleft}
\footnotesize
$t_0$ is in $\mathrm{MeV\,fm^3}$;
$t_1,t_2,W_0$ are in $\mathrm{MeV\,fm^5}$; and
$t_3$ is in $\mathrm{MeV\,fm^{3+3\alpha}}$.
\end{flushleft}
\end{table*}

\section{Local sensitivity coefficients, fit definition, and validation}
\label{sec:supp_sensitivity}

The surface convention used in the code obeys
\begin{equation}
 \rho_0\Delta\rho_1+\rho_1\Delta\rho_0
 =2\left(\rho_n\Delta\rho_n-\rho_p\Delta\rho_p\right),
 \label{eq:supp_surface}
\end{equation}
with $\rho_0=\rho_n+\rho_p$ and $\rho_1=\rho_n-\rho_p$.
After integration by parts, Eq.~(\ref{eq:supp_surface}) is proportional
to $-2[(\nabla\rho_n)^2-(\nabla\rho_p)^2]$.  This identity fixes the
sign and normalization of $C_\Delta^{\rm III}$.

For each EDF and mirror pair, the five coupling configurations used to build the local sensitivities are listed in Table~\ref{tab:supp_coupling_points}.  The same value of the chosen coupling is applied to both mirror members; the mirror differences are formed only after the two self-consistent calculations are completed.

\begin{table}[t]
\centering
\caption{Coupling configurations used to construct the local sensitivity matrix for each EDF and mirror pair.}
\label{tab:supp_coupling_points}
\small
\begin{tabular}{lccp{0.36\columnwidth}}
\toprule
Configuration & $t_0^{\rm III}$ & $C_\Delta^{\rm III}$ & Purpose \\
\midrule
baseline & 0 & 0 & CSB-free reference value \\
$t_0+$ & $+h_0$ & 0 & positive volume-CSB variation \\
$t_0-$ & $-h_0$ & 0 & negative volume-CSB variation \\
$C_\Delta+$ & 0 & $+h_\Delta$ & positive surface-gradient variation \\
$C_\Delta-$ & 0 & $-h_\Delta$ & negative surface-gradient variation \\
\bottomrule
\end{tabular}
\end{table}

For an observable
$\mathcal O_i\in\{\Delta B_i,\Delta R_{{\rm ch},i}^{\rm mirr},
\Delta R_{np,i}^{p}\}$, the self-consistent local slopes are
\begin{align}
 S_{0,i}^{\mathcal O}
 &=\frac{\mathcal O_i(+h_0,0)-\mathcal O_i(-h_0,0)}{2h_0},
 \label{eq:supp_s0}\\
 S_{\Delta,i}^{\mathcal O}
 &=\frac{\mathcal O_i(0,+h_\Delta)-\mathcal O_i(0,-h_\Delta)}
 {2h_\Delta}.
 \label{eq:supp_sd}
\end{align}
The values at fitted couplings are obtained from the linearized
model
\begin{equation}
 \mathcal O_i(t_0^{\rm III},C_\Delta^{\rm III})
 \simeq \mathcal O_i^{(C)}
 +S_{0,i}^{\mathcal O}t_0^{\rm III}
 +S_{\Delta,i}^{\mathcal O}C_\Delta^{\rm III}.
 \label{eq:supp_linear}
\end{equation}
For the MDE and mirror charge-radius difference, these slopes form the local sensitivity matrix
\begin{equation}
 {\bf S}_i=
 \begin{pmatrix}
 S_{0,i}^{B} & S_{\Delta,i}^{B}\\
 S_{0,i}^{R} & S_{\Delta,i}^{R}
 \end{pmatrix},
 \label{eq:supp_Smatrix}
\end{equation}
which satisfies
\begin{equation}
 \begin{pmatrix}
  \delta_{\rm CSB}\Delta B_i\\[2pt]
  \delta_{\rm CSB}\Delta R_{{\rm ch},i}^{\rm mirr}
 \end{pmatrix}
 \simeq
 {\bf S}_i
 \begin{pmatrix}
  t_0^{\rm III}\\[2pt]
  C_\Delta^{\rm III}
 \end{pmatrix}.
 \label{eq:supp_matrix_response}
\end{equation}
The diagnostic ratios are
\begin{equation}
 \lambda_i^B=-\frac{S_{\Delta,i}^{B}}{S_{0,i}^{B}},
 \qquad
 \lambda_i^R=-\frac{S_{\Delta,i}^{R}}{S_{0,i}^{R}}.
 \label{eq:supp_lambdas}
\end{equation}
The ratios display the MDE and mirror-radius directions in the
$(t_0^{\rm III},C_\Delta^{\rm III})$ coupling plane, but they are not
fitting inputs.  The full mass-radius fit uses the four coefficients in
Eq.~\eqref{eq:supp_Smatrix}.  The full production values for the four
measured calibration pairs are collected in
Table~\ref{tab:local_sensitivity_matrix}, which is the numerical
counterpart of Eqs.~\eqref{eq:supp_Smatrix}--\eqref{eq:supp_lambdas}.

\begin{table*}[t]
\centering

\caption{
Full local sensitivity matrix for the four measured calibration pairs.
For each EDF and mirror pair $i$, the entries
$S_{0,i}^{B}$, $S_{\Delta,i}^{B}$, $S_{0,i}^{R}$, and
$S_{\Delta,i}^{R}$ are the production finite-difference slopes entering
Eq.~\eqref{eq:supp_Smatrix}. The ratios
$\lambda_i^B=-S_{\Delta,i}^{B}/S_{0,i}^{B}$ and
$\lambda_i^R=-S_{\Delta,i}^{R}/S_{0,i}^{R}$ are diagnostic quantities, not fitting inputs.
The same slopes form the design matrix of the full mass--radius fit.
}
\label{tab:local_sensitivity_matrix}
\renewcommand{\arraystretch}{1.15}
\setlength{\tabcolsep}{5pt}
\scriptsize
\resizebox{\textwidth}{!}{%
\begin{tabular}{llcccccc}
\hline
EDF & Pair
& \(S_{0,i}^{B}\)
& \(S_{\Delta,i}^{B}\)
& \(\lambda_i^{B}\)
& \(S_{0,i}^{R}\)
& \(S_{\Delta,i}^{R}\)
& \(\lambda_i^{R}\) \\
&
& \([\mathrm{fm}^{-3}]\)
& \([\mathrm{fm}^{-5}]\)
& \([\mathrm{fm}^{-2}]\)
& \([\mathrm{MeV}^{-1}\mathrm{fm}^{-2}]\)
& \([\mathrm{MeV}^{-1}\mathrm{fm}^{-4}]\)
& \([\mathrm{fm}^{-2}]\) \\
\hline
SLy4
& \(^{34}\mathrm{Ar}\)--\(^{34}\mathrm{S}\)
& \(+0.1457\) & \(-0.1860\) & \(1.2763\)
& \(-2.7777\!\times\!10^{-4}\) & \(+6.5450\!\times\!10^{-4}\) & \(2.3563\) \\
& \(^{36}\mathrm{Ca}\)--\(^{36}\mathrm{S}\)
& \(+0.2993\) & \(-0.3875\) & \(1.2949\)
& \(-5.7884\!\times\!10^{-4}\) & \(+1.3760\!\times\!10^{-3}\) & \(2.3772\) \\
& \(^{38}\mathrm{Ca}\)--\(^{38}\mathrm{Ar}\)
& \(+0.1531\) & \(-0.2056\) & \(1.3436\)
& \(-2.8839\!\times\!10^{-4}\) & \(+6.9000\!\times\!10^{-4}\) & \(2.3926\) \\
& \(^{54}\mathrm{Ni}\)--\(^{54}\mathrm{Fe}\)
& \(+0.1750\) & \(-0.2379\) & \(1.3589\)
& \(-2.3054\!\times\!10^{-4}\) & \(+5.0250\!\times\!10^{-4}\) & \(2.1797\) \\
\hline
SkM*
& \(^{34}\mathrm{Ar}\)--\(^{34}\mathrm{S}\)
& \(+0.14211\) & \(-0.1686\) & \(1.1860\)
& \(-2.7348\!\times\!10^{-4}\) & \(+5.6050\!\times\!10^{-4}\) & \(2.0494\) \\
& \(^{36}\mathrm{Ca}\)--\(^{36}\mathrm{S}\)
& \(+0.2867\) & \(-0.3285\) & \(1.1458\)
& \(-5.9768\!\times\!10^{-4}\) & \(+1.2675\!\times\!10^{-3}\) & \(2.1207\) \\
& \(^{38}\mathrm{Ca}\)--\(^{38}\mathrm{Ar}\)
& \(+0.1468\) & \(-0.1752\) & \(1.1938\)
& \(-2.8571\!\times\!10^{-4}\) & \(+5.7250\!\times\!10^{-4}\) & \(2.0038\) \\
& \(^{54}\mathrm{Ni}\)--\(^{54}\mathrm{Fe}\)
& \(+0.1728\) & \(-0.2203\) & \(1.2749\)
& \(-2.4277\!\times\!10^{-4}\) & \(+4.5150\!\times\!10^{-4}\) & \(1.8598\) \\
\hline
\end{tabular}%
}
\end{table*}

The first row of Table~\ref{tab:local_sensitivity_matrix} is the SLy4
$^{34}$Ar--$^{34}$S example quoted in the main text.  It gives
$\lambda^B=1.2763~\mathrm{fm^{-2}}$ and
$\lambda^R=2.3563~\mathrm{fm^{-2}}$, illustrating how the tabulated slopes
are converted into the diagnostic ratios.  The main text discusses the
physical interpretation; here the table is provided as the numerical input
for the linearized calibration.

Thus the direct HFB calculations determine local self-consistent slopes,
whereas the reported fitted residuals and predictions are
linearized extrapolations from those slopes. Here “odd-in-coupling” means the component of the observable change that changes sign under \(g\to -g\). 
The residual non-odd component is the even finite-step remainder,
\[
\epsilon_i^B(h)
=
\frac{\Delta B_i(+h)+\Delta B_i(-h)-2\Delta B_i(0)}{2},
\]
which would vanish for an exactly linear dependence.

The explicit CSB-energy contribution is odd under coupling reversal at
approximately the $10^{-3}$ MeV level.  For the total self-consistent
energy shift, the largest non-odd contribution in the production tests
is about $0.026$ MeV, below the adopted $0.05$ MeV validation
criterion.  A diagnostic local electromagnetic $pp$ contact term,
varied over $-0.10\le C_{\rm EM}^{pp}\le0.10~\mathrm{MeV\,fm^3}$,
produces only small changes in the fit diagnostics and is not included
in the production parameter set. The \(C_{\rm EM}^{pp}\) scan is a schematic proton-only zero-range
diagnostic, formally analogous to the \(pp\) component of a Skyrme
contact term. It is not part of the standard Skyrme functional used in
the production calculations.

Using the residuals defined in the main text, the fitted couplings
minimize
\begin{equation}
 \chi^2=
 \sum_i\left(\frac{R_{B,i}^{(C+{\rm CSB})}}{\sigma_{B,i}}\right)^2
 +\eta_R
 \sum_i\left(\frac{R_{R,i}^{(C+{\rm CSB})}}{\sigma_{R,i}}\right)^2,
 \label{eq:supp_chi2}
\end{equation}
where
\begin{align}
\sigma_{B,i}^2 &= (\sigma_{B,i}^{\rm exp})^2 + (0.10~\mathrm{MeV})^2, \nonumber \\
\sigma_{R,i}^2 &= (\sigma_{R,i}^{\rm exp})^2 + (0.005~\mathrm{fm})^2.
\label{eq:supp_errors}
\end{align}
Here $\eta_R=0$ denotes the MDE-only fit and $\eta_R=1$ the full
mass--radius fit.  For the full fit, the four measured calibration pairs define the design matrix
\begin{equation}
 X=
 \begin{pmatrix}
 S_{0,1}^{B} & S_{\Delta,1}^{B}\\
 \vdots & \vdots\\
 S_{0,4}^{B} & S_{\Delta,4}^{B}\\
 S_{0,1}^{R} & S_{\Delta,1}^{R}\\
 \vdots & \vdots\\
 S_{0,4}^{R} & S_{\Delta,4}^{R}
 \end{pmatrix},
 \qquad
 r=
 \begin{pmatrix}
 R_{B,1}^{(C)}\\ \vdots\\ R_{B,4}^{(C)}\\
 R_{R,1}^{(C)}\\ \vdots\\ R_{R,4}^{(C)}
 \end{pmatrix}.
 \label{eq:supp_design_matrix}
\end{equation}
With the weight matrix $W$, the fitted coupling vector
$g=(t_0^{\rm III},C_\Delta^{\rm III})^T$ is
\begin{equation}
 g_{\rm fit}=(X^T W X)^{-1}X^T W r.
 \label{eq:supp_weighted_fit}
\end{equation}
Thus the full fit uses the full sensitivity matrix, not the compressed ratios $\lambda_i^B$ or $\lambda_i^R$ alone. The entries of \(X\) are precisely the production slopes listed in
Table~\ref{tab:local_sensitivity_matrix}. Thus Table~\ref{tab:local_sensitivity_matrix} provides the numerical input for the linearized
full mass--radius fit, whereas the ratios \(\lambda_i^B\) and
\(\lambda_i^R\) are only post-processing diagnostics of the same matrix.

The covariance matrix is the inverse weighted normal
matrix of this linear least-squares problem, and the quoted coupling
errors are the square roots of its diagonal elements.  The RMS values in
Table~1 of the main text are unweighted RMS residuals over the four
measured calibration pairs, while $\chi^2/{\rm dof}$ uses
${\rm dof}=N_{\rm data}-N_{\rm par}$.  The strong positive parameter
covariance reflects compensation between the two MDE contributions
because $S_0^B$ and $S_\Delta^B$ have opposite signs.

The sensitivity ratio and its calibration-pair average are
\begin{equation}
 \lambda_i^B=-\frac{S_{\Delta,i}^B}{S_{0,i}^B},\qquad
 \lambda_{\rm cl}=\frac{1}{4}\sum_{i=1}^{4}\lambda_i^B.
 \label{eq:supp_lambda}
\end{equation}
The quoted uncertainty of $\lambda_{\rm cl}$ is the sample standard
deviation across the four measured calibration pairs, not a statistical fit error.  The analogous radius ratios $\lambda_i^R$ are computed from the same production slopes.  

\section{Calibration and construction of the predictions}
\label{sec:supp_prediction}

The four measured calibration pairs are
$^{34}$Ar--$^{34}$S, $^{36}$Ca--$^{36}$S,
$^{38}$Ca--$^{38}$Ar, and $^{54}$Ni--$^{54}$Fe.  Their sensitivity-ratio
averages are
\begin{align}
 \lambda_{\rm cl}^{\rm SLy4}
 &=1.3185\pm0.0391~\mathrm{fm^{-2}},\\
 \lambda_{\rm cl}^{\rm SkM^*}
 &=1.2001\pm0.0541~\mathrm{fm^{-2}}.
 \label{eq:supp_lambda_values}
\end{align}

The physical interpretation of the MDE-only and full mass-radius calibrations,
including the residual comparison shown in Fig.~2 of the main text, is
given in the main article.  For reproducibility, we list here the
full-precision full-fit couplings used in the target predictions:
\begin{align}
 (t_0^{\rm III},C_\Delta^{\rm III})_{\rm SLy4}
 &=(-25.8204,-14.7126),\\
 (t_0^{\rm III},C_\Delta^{\rm III})_{\rm SkM^*}
 &=(-24.7262,-15.2539),
 \label{eq:supp_joint_values}
\end{align}
where the first and second entries are in
$\mathrm{MeV\,fm^3}$ and $\mathrm{MeV\,fm^5}$, respectively.  These
values give
\begin{align}
 t_{\rm eff}^{\rm III}({\rm SLy4})
 &=t_0^{\rm III}-\lambda_{\rm cl}C_\Delta^{\rm III}
 =-6.4226~\mathrm{MeV\,fm^3},\\
 t_{\rm eff}^{\rm III}({\rm SkM^*})
 &=t_0^{\rm III}-\lambda_{\rm cl}C_\Delta^{\rm III}
 =-6.4196~\mathrm{MeV\,fm^3}.
 \label{eq:supp_teff_values}
\end{align}
Neither target-pair MDEs nor proton-rich target radii enter this
calibration.  For each target pair, the same finite-difference construction
is used to obtain target-specific radius slopes.

For a target pair $j$,
\begin{align}
 \Delta R_{{\rm ch},j}^{\rm mirr,pred}
 &=\Delta R_{{\rm ch},j}^{\rm mirr,(C)}
 +S_{0,j}^{R}t_0^{\rm III}
 +S_{\Delta,j}^{R}C_\Delta^{\rm III},
 \label{eq:supp_pred_diff}\\
 R_{{\rm ch},j}^{p,\rm pred}
 &=R_{{\rm ch},j}^{n,\rm exp}
 +\Delta R_{{\rm ch},j}^{\rm mirr,pred}.
 \label{eq:supp_pred_abs}
\end{align}
The parenthetical uncertainty in Table~3 of the main text is
\begin{equation}
 \sigma_{{\rm min},j}
 =\sqrt{(\sigma_{{\rm exp},j}^{n})^2+(0.005~\mathrm{fm})^2},
 \label{eq:supp_sigma_min}
\end{equation}
and the separately quoted two-EDF difference is
\begin{equation}
 \delta_{{\rm EDF},j}
 =\left|\Delta R_{{\rm ch},j}^{\rm SLy4}
 -\Delta R_{{\rm ch},j}^{\rm SkM^*}\right|.
 \label{eq:supp_delta_edf}
\end{equation}
The latter is a limited diagnostic of model dependence, not a
statistical standard deviation.  Coupling-covariance and
$\lambda_B$-band assignment uncertainties are not included in the parenthetical
errors.

The prediction pairs are compatible with the compact calibration-pair $\lambda_B$ band, although
their mean $\lambda_B$ is lower than the calibration-pair mean in SLy4.  As
low-$\lambda_B$ diagnostic checks,
$^{30}$S--$^{30}$Si gives $\lambda_B=0.4176$ (SLy4) and
$0.6828~\mathrm{fm^{-2}}$ (SkM*), while the measured
$^{32}$Ar--$^{32}$Si pair~\cite{Konig2024Silicon} gives $0.8741$ and
$0.9300~\mathrm{fm^{-2}}$, respectively.

\FloatBarrier